\documentclass[manuscript,screen]{acmart}

\AtBeginDocument{%
  }

\begin{document}

\title{Mental-Gen: A Brain-Computer Interface-Based Interactive Method for Interior Space Generative Design}

\author{YIJIANG LIU}
\affiliation{%
  \institution{Tsinghua University}
  \city{Haidian Qu}
  \state{Beijing Shi}
  \country{China}}
\email{liuyijia19@mails.tsinghua.edu.cn}

\author{HUI WANG}
\affiliation{%
  \institution{Tsinghua University}
  \city{Haidian Qu}
  \state{Beijing Shi}
  \country{China}}
\email{wh-sa@mail.tsinghua.edu.cn}


\begin{abstract}
Interior space design significantly influences residents' daily lives. However, the process often presents high barriers and complex reasoning for users, leading to semantic losses in articulating comprehensive requirements and communicating them to designers.
This study proposes the Mental-Gen design method, which focuses on interpreting users' spatial design intentions at neural level and expressing them through generative AI models. We employed unsupervised learning methods to detect similarities in users' brainwave responses to different spatial features, assess the feasibility of BCI commands. We trained and refined generative AI models for each valuable design command. The command prediction process adopted the motor imagery paradigm from BCI research. We trained Support Vector Machine (SVM) models to predict design commands for different spatial features based on EEG features. The results indicate that the Mental-Gen method can effectively interpret design intentions through brainwave signals, assisting users in achieving satisfactory interior space designs using imagined commands.
\end{abstract}




\maketitle

\section{INTRODUCTION}
As technology and productivity continue to advance, the ability to meet increasingly customized demands has become a reality, leading to more diverse design challenges and scenarios in people's daily lives. However, ordinary users are often untrained in design expression and face difficulties using high-barrier tools like drawing and modeling to accurately convey their design ideas. Moreover, abstract expressions through language and text often result in information loss. For instance, in the context of home interior design, each user possesses unique design needs. Still, due to their own challenges in design, expression, and communication, they often end up with mediocre, homogeneous, and unsatisfactory design outcomes.

The emergence of generative artificial intelligence tools in recent years has partially alleviated this dilemma. AI can assist in expressing users' design ideas to some extent. Take the Stable Diffusion platform as an example: by inputting simple semantic prompts and possible composition references, users can achieve a direct visual representation of their ideas and receive detailed architectural enhancements. However, while AI-generated architectural images are becoming increasingly realistic and physically coherent as the technology evolves, they often exhibit ambiguity and uncertainty in the realization of specific design details. The same semantic prompt may lead to various image characteristics in different contexts, making it challenging for users to accurately express their design intentions. On the other hand, even with sufficient design expression experience, users' design ideas are often unclear in the initial stages and typically require ongoing self-innovation and refinement during the expression process. Therefore, real-time emotions and subconscious thoughts significantly influence the final design outcome.

The development of neuroscience offers a promising solution to these issues. With advancements in understanding the human brain's imagination mechanisms and breakthroughs in decoding algorithms, we can now use machine learning models to analyze the temporal and frequency domain characteristics of brain electrical signals, allowing for the decoding and categorization of some of the brain's ideas. This makes it possible to interpret users' design intentions in specific domains and dimensions using brain-computer interface (BCI) technology.

Consequently, this study proposes a human-computer interaction paradigm that combines brain-computer interface (BCI) technology with generative artificial intelligence (Gen-AI). The goal is to decode users' design ideas from their brain signals and enable precise interactive spatial design through the synergy of users' imaginative intentions and the expressive capabilities of generative AI.
\begin{figure}[h]
  \centering
  \includegraphics[width=\linewidth]{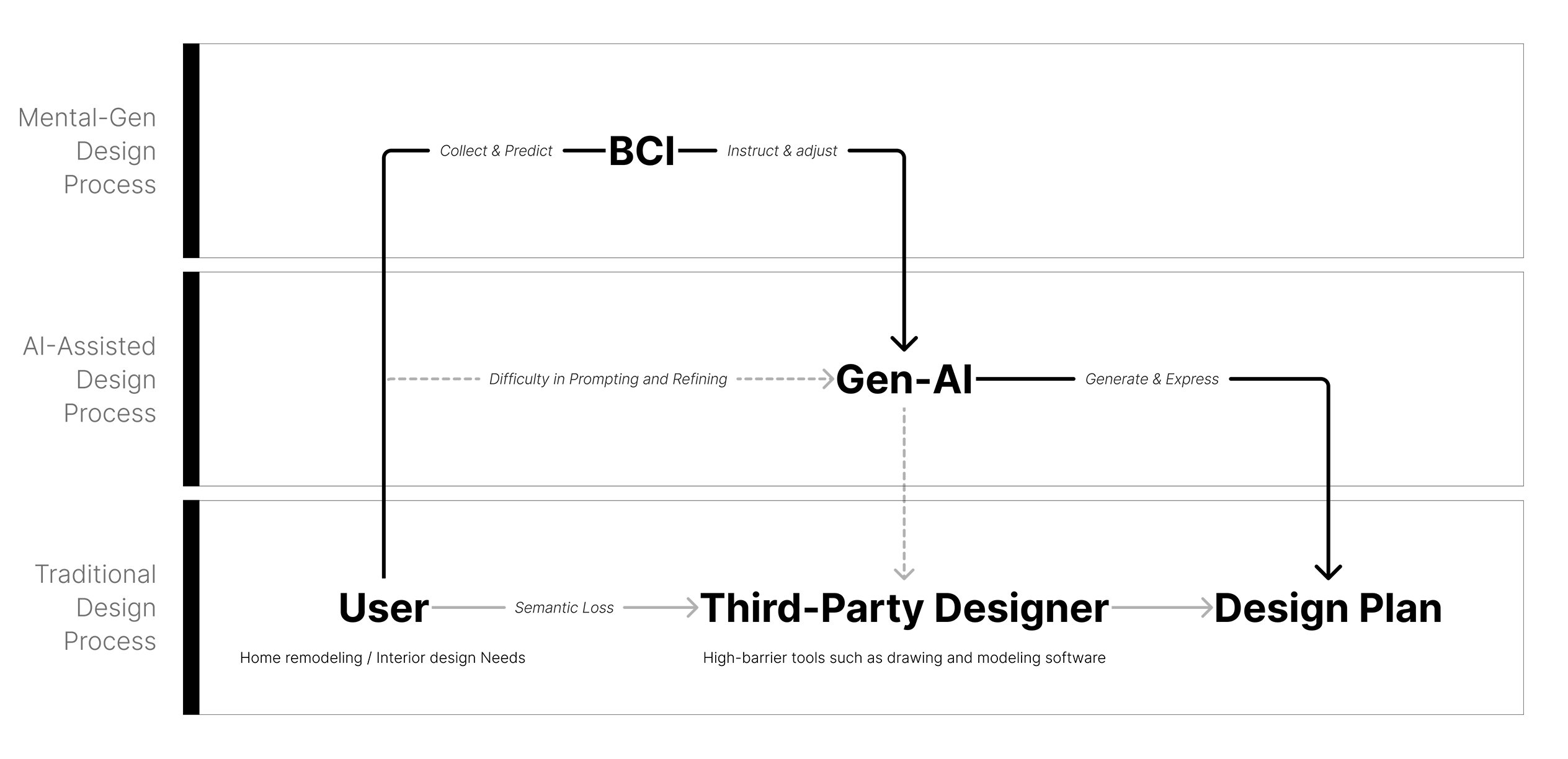}
  \caption{Mental-Gen Method Compared to Traditional Design Processes}
  \Description{Mental-Gen Method Compared to Traditional Design Processes}
\end{figure}

\section{RELATED WORK}

\subsection{Cognitive States in Design Process}
Beyond practical challenges, such as users' unfamiliarity with design tools, an ideal design process does not simply involve translating a preconceived and refined idea into a final product. Instead, it is a continuous process of iteration, where expression and creative cognition evolve to achieve a well-developed design outcome. Finke proposed that human creative cognition consists of two stages: generation and exploration. The generation stage leads to the creation of preinventive structures, which are holistic and broad imaginings of the design outcome. The subsequent exploration stage is driven by cognitive and experiential actions, gradually refining the preinventive structures toward completion \cite{art1}. Building on this, Taura et al. further categorized design behavior into low-level concepts driven by problem-solving and high-level concepts driven by intrinsic feelings and subconscious cognition \cite{art2}. In spatial design, the expression of concrete and operational design ideas can often be broken down into discrete steps and is relatively straightforward to implement, such as the placement of windows, the angles of lines, or the arrangement of furniture. However, conceptualizing and expressing a comprehensive and complex design vision, which encompasses an exponential number of individual design ideas, is often more challenging. This process requires a nuanced understanding of the overall complexity, which is the primary focus of this study.

Focusing on interior space design, where the basic spatial structure is already defined, the most critical factors for users to consider are style \cite{art3}, material, and color coordination \cite{art4}. According to the annual home design trend report by Houzz, a leading media platform in the interior design industry, style and functionality are the primary drivers of home renovations. "Style" in this context can be deconstructed into two main aspects: firstly, it represents an aesthetic form and expression method reflective of a specific cultural background, such as modern, Scandinavian, or new Chinese styles. Secondly, it refers to the visual expressiveness and complexity of the design, such as minimalist, casual luxury, or high luxury styles. Additionally, while material coordination typically falls under style design, the balance and combination of solid and transparent materials require independent consideration. The openness and transparency of interior design significantly affect a person’s visual experience, subsequently influencing their spatial experience and even their level of spatial cognition \cite{art5}.

\subsection{Generative AI for Design Process}
Early generative artificial intelligence models primarily originated from probabilistic models such as Naive Bayes and Hidden Markov Models, which were used to model the probability distribution of data. With the advancement of deep learning techniques, more expressive deep generative models like Generative Adversarial Networks (GANs) and Variational Autoencoders (VAEs) have gradually become mainstream. More recently, the diffusion model, introduced by Sohl-Dickstein et al., generates data by iteratively diffusing an initial Gaussian noise vector, gradually approximating the target data distribution. This allows for the imitation of known images in terms of style or information. Building upon diffusion models, further developments like Latent Diffusion Models and Stable Diffusion models have achieved more sophisticated image generation capabilities, enabling a certain degree of semantic alignment with input prompts. By this stage, the foundational technologies for image generation using generative models have become highly advanced.

Furthermore, researchers have extended the capabilities of Stable Diffusion models by developing autoregressive generation models such as the LoRA (Logarithmically-Decaying Recurrent Attention) model and the ControlNet model, which incorporates control mechanisms. The LoRA model serves as a pre-trained style fine-tuning model designed to better control the stylistic effects of the generated images. Meanwhile, the control mechanism in the ControlNet model allows users to influence the features of the generated images during the creation process, enhancing the precision of generative outputs. In terms of the application of generative models, researchers have integrated various machine learning techniques such as masking \cite{art6}, semantic segmentation \cite{art7}, and cross-attention \cite{art8} to improve the utility of generative models across different specific domains.

While generative AI, through the use of prompts and model training, can assist non-professional designers in the general public in initially expressing their spatial design intentions, it still suffers from significant randomness and ambiguity. The relationship between prompts and design outcomes remains overly abstract, making it challenging for users to achieve refined design expression comprehensively \cite{art9, art10}.

\subsection{BCI: Development and Design Applications}
Electroencephalography (EEG) is a crucial tool in neuroscience research used to record electrical activity in the brain. It was first discovered and recorded by Hans Berger in 1924. Modern EEG measurements primarily involve placing electrodes on the scalp to capture the electric fields generated by the synchronous activity of large groups of neurons in the cerebral cortex. EEG signals typically represent microvolt-level potential changes and include oscillations across various frequency bands, such as delta ($\delta$, 0.5-4 Hz), theta ($\theta$, 4-8 Hz), alpha ($\alpha$, 8-13 Hz), beta ($\beta$, 13-30 Hz), and gamma ($\gamma$, 30-100 Hz) waves. These oscillations provide detailed insights into the electrophysiological changes occurring during brain activity.

In practical research, feature extraction methods are often employed to transform high-dimensional EEG signals into lower-dimensional feature data to analyze their correlation with specific research questions. For temporal domain features, mean and variance are commonly used to describe the overall activity level of EEG signals, while amplitude and power represent the potential strength and energy of EEG signals over different time periods. In the frequency domain, researchers often use band power to differentiate various brain states and apply Fast Fourier Transform (FFT) to compute the Power Spectral Density (PSD) of EEG signals to analyze the energy distribution across different frequency bands \cite{art11}.

As research has progressed, classical paradigms such as Motor Imagery (MI) and Event-Related Potentials (ERPs) have been developed within the EEG field. Motor Imagery involves recognizing specific brain region activities by imagining particular movements without actual physical actions, allowing for the classification and prediction of EEG-based intentions \cite{art12}. Traditional motor imagery methods include Common Spatial Pattern (CSP) and Linear Discriminant Analysis (LDA). However, with the maturation of machine learning techniques, researchers have begun using Convolutional Neural Networks (CNNs) and Recurrent Neural Networks (RNNs) to directly extract features from raw EEG data for classification. Among these methods, Support Vector Machine (SVM) models, which use hyperplanes to distinguish between different feature vectors, have been shown to achieve high accuracy in EEG classification \cite{art13}.

Regarding the application of EEG in design control, most existing studies adopt an approach that involves mapping user EEG affective data, event-related potentials, and other indicators to design parameters, focusing primarily on the indirect mapping of EEG intentions \cite{art14, art15, art16, art17}. In the direct design control process using EEG-based machine learning predictions, the complexity and diversity of design problems often present the most significant challenges to be addressed. For example, Qi et al. decomposed complex design problems into several simple geometric shapes, guiding participants to observe changes in these shapes while collecting their EEG data. This approach enabled the identification of corresponding geometric shapes and their weights from the participants' imagined intentions, allowing for the weighted combination of these shapes to output a design result through brain-computer interaction \cite{art18}. Although the EEG imagination mechanisms associated with complex design problems are challenging to interpret using current technological methods, researchers have been able to identify EEG patterns corresponding to specific dimensions of design problems through deconstruction and analysis. This has enabled the use of brain-computer interface systems to predict and express design intentions.

\section{Design Command Study}
To identify which spatial features and design dimensions users are more concerned with in real-world interior design scenarios, the study first evaluated various types of spatial transformations from two perspectives: value and feasibility. This approach aimed to determine the most appropriate brain-computer interface (BCI) design commands.

\subsection{User Interviews}
First, we conducted user interviews to preliminarily identify what types of design adjustments are most needed. We recruited a total of 12 participants, including 7 students from non-design disciplines and 5 online recruits who currently have real home renovation needs. The participants ranged in age from 20 to 37 years old. They were asked to consider their primary needs in interior design from the perspective of the design, decoration, and renovation of their own residences. The researchers then performed content analysis and thematic modeling to determine which spatial features of interior design users value the most.

The interview results indicated that participants often prioritize overall spatial style, color, lighting, and transparency as their primary considerations. Typical responses included statements like, "I want my home to lean more towards a Chinese style with a touch of understated luxury, with good lighting but without sacrificing privacy." While many users also emphasized the importance of functional features such as accessibility and comfort, they also acknowledged that these features are more significantly influenced by external factors like floor plans. Therefore, the researchers conducted text frequency analysis and thematic modeling on the filtered interview content to establish that the primary focus of interior space design needs in this study centers around four spatial features: "transparency degree, style preference, decoration density, and color scheme." Consequently, design commands in the interactive system that can adjust these four characteristics would have greater practical value.

The "Transparency degree" emphasizes whether the space is more private, conservative, and enclosed or more open, transparent, and expansive. "Style preference" focuses on whether the interior design leans more toward modernism or a more classical aesthetic. From the interview content, "classical" in the context of this study specifically refers to traditional Chinese architectural styles. "Decoration density" represents whether users prefer a more minimalist and clean design or a more luxurious and ornate decoration style. Finally, "Color scheme" encompasses a range of palettes, from cool, low-saturation color schemes to bold, bright, and vivid ones, as well as all the variations in between. The study summarized two design directions encompassed by these spatial characteristics, as shown in Table~\ref{tab1}.

\begin{table*}
  \caption{Selection of Spatial Features}
  \label{tab1}
  \begin{tabular}{ccccc}
    \toprule
    Type&Transparency Degree&Style Preference&Decoration Density&Color Scheme\\
    \midrule
    Defined positive(+) & open, transparent& classicism&luxurious,  ornate&bright, colorful\\
    Defined negative(-) & private, conservative & modernism&minimalist, clean&cool, monotonous\\
  \bottomrule
\end{tabular}
\end{table*}

\subsection{Exploratory Experiment}
To validate whether the aforementioned spatial features can elicit similar brain activity patterns, thereby determining the feasibility of effectively training design commands to modify these features, the researchers designed a "viewing-imagination" experiment for each spatial feature variation and performed unsupervised clustering on the collected EEG data.

The research team collected 410 high-quality interior design images from mainstream design websites such as Pinterest and Houzz. These images were then evaluated by experts based on the four spatial features previously identified: transparency degree, style preference, decoration density, and color scheme. The goal was to determine if each image prominently displayed any specific spatial feature. The score range for each feature was defined from -5 to 5, where 0 represented a neutral expression, -5 indicated an extreme expression in the opposite direction (e.g., highly private, modern style, extremely minimalist, cool/dark color scheme), and 5 represented an extreme positive expression (e.g., highly open, classical style, highly luxurious, saturated color scheme). After three experts rated the images, the average scores for each of the four features were calculated. Images that did not exhibit any significant feature were removed, and overrepresented features, such as extreme openness, were filtered and reduced. This process resulted in a final dataset of 84 images, ensuring a balanced representation of each spatial feature in both positive and negative directions.

For each interior design image, a white model base image was created, along with a transition video from the base image to the design image, to aid participants in their imagination tasks. Each video was 8 seconds long, including a 3-second transition that repeated twice, with a 2-second black screen frame inserted between each video for seamless switching to the next one.

We recruited 15 participants, all of whom were university students with normal vision, aged between 19 and 24 years. Their primary task was to watch the videos and imagine the spatial features manifesting according to their subjective intentions. Meanwhile, their EEG data were recorded. The study utilized the EPOC X device developed by EMOTIV, which features 14 channels and adheres to the international 10-20 system for electrode placement, with a signal sampling rate of 256 Hz. Additionally, the study employed the EMOTIV PRO Standard license and BCI-OSC plugin for the collection and real-time transmission of raw EEG signals. Each participant's EEG data were segmented according to the video timestamps, resulting in 84 segments corresponding to the 84 interior design images, each annotated with the four spatial feature scores.
\begin{figure}[h]
  \centering
  \includegraphics[width=\linewidth]{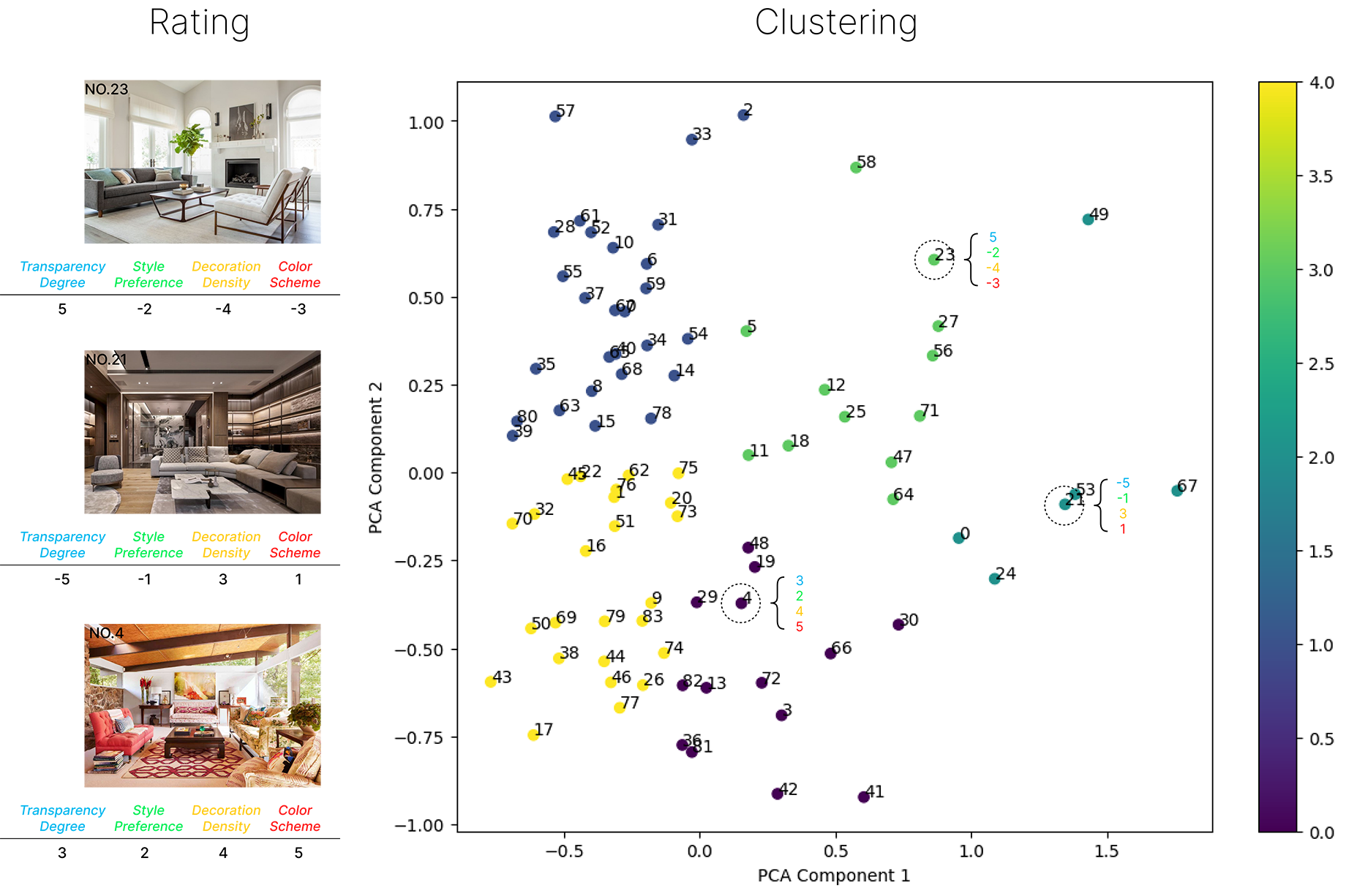}
  \caption{Image Labeling and Corresponding EEG Data Clustering}
\end{figure}

\subsection{Data Processing and EEG Clustering}
The experiment collected EEG data from participants as they watched videos and imagined various spatial feature changes. Through preprocessing, dimensionality reduction, and unsupervised clustering algorithms, we aimed to investigate whether the EEG data corresponding to the imagination of the same spatial feature changes, under different conditions, exhibit commonalities. This would indicate whether these changes are suitable for forming commands in a brain-computer interface (BCI) system.

The preprocessing of EEG signals and the unsupervised clustering process were implemented in the online coding environment Colab. The EEG signal processing primarily utilized the MNE standard library, while K-means clustering was performed using the sklearn library in Python. The specific process was as follows: First, the collected raw EEG data underwent filtering and Independent Component Analysis (ICA) to remove artifacts, retaining brainwave frequencies in the range of 0.5 Hz to 45 Hz. The Welch method was then used to analyze the power distribution of each frequency component in the EEG signals, resulting in the corresponding power spectral density (PSD) of the EEG. Subsequently, Principal Component Analysis (PCA) was applied to reduce the dimensionality of the high-dimensional PSD feature data to a two-dimensional space, identifying the principal components of the data to make subsequent clustering more efficient. Finally, the processed EEG segments were clustered using the K-means algorithm.

\subsection{Results Analysis and Command Optimization}
Based on the dual evaluation metrics of silhouette coefficient and Calinski-Harabasz Index, the K-means clustering achieved the best performance when the data was partitioned into five clusters. Consequently, the study adopted this five-cluster configuration and analyzed the relationship between the clustering results of the EEG data and the spatial feature scores associated with each cluster. Given that we evaluated spatial features across four dimensions, each segment of EEG data was tagged with four labels, each corresponding to only two directions (positive or negative) and their respective weights. This significantly reduced the potential for cluster skewness and class imbalance issues. Additionally, due to the complexity of EEG data and the potential for random consistency, it is challenging to use common evaluation methods such as Rand Index (RI) or Fowlkes-Mallows Index (FMI) to assess the clustering results. Therefore, the study opted to use purity to analyze the impact of each spatial feature on the clustering outcomes.

The calculation of purity typically involves recording the most frequent label within each cluster, summing the occurrences of this representative label across all clusters, and examining its ratio to the total dataset size. In this study, given that each of the four spatial features can have either a positive or negative label along with a corresponding score indicating its magnitude, the researchers considered the score as the label's weight (with scores ranging from 0 to 5 normalized to a weight range of 0 to 1, where each unit corresponds to a weight increment of 0.2). The representative label within each cluster was determined by multiplying the occurrence frequency of each direction (positive or negative) by its corresponding weight. Similarly, the final purity value also needed to account for the influence of these weights; it was calculated by multiplying the count of representative labels in each cluster by their weights, summing these products, and then dividing by the sum of all data points multiplied by their respective weights. 

For a specific spatial feature, assume the total number of data samples is \( N \), in which each sample \( i \) has its label \( T_j \) and a corresponding weight \( w_i \). Samples are categorized according to clusters \( C_i \), the weighted purity can be calculated using the following formula:

\[
\text{Weighted Purity} = \frac{1}{W} \sum_{i=1}^{k} \left( \sum_{x \in C_i \cap L_{j^*}} \left| w_x \right| \right)
\]

where:

\begin{itemize}
    \item \( W \) is the sum of the absolute weights of all \( N \) samples 
    \item \( C_i \) is cluster \( i \).
    \item \( L_{j^*} \) is the label with the maximum weight in cluster \( C_i \).
    \item \( w_x \) is the weight of sample \( x \).
    \item \( \sum_{x \in C_i \cap L_{j^*}} \left| w_x \right| \) is the sum of the absolute weights values of all samples in cluster \( C_i \) that belong to label \( L_{j^*} \).
\end{itemize}

In addition to purity, the researchers also analyzed the V-Measure of the clustering results. V-Measure is the harmonic mean of homogeneity and completeness in clustering results, aimed at balancing these two aspects. Homogeneity indicates that all data points within a cluster should ideally have the same label, while completeness ensures that all data points with the same label are ideally grouped within the same cluster. The former is similar to the function of purity in evaluating clustering quality, while the latter compensates for any limitations of the purity metric.

We recorded the clustering results for all 84 segments of EEG data from each participant. For each spatial feature, we calculated the purity based on the data labels and weights for that feature and subsequently computed the V-Measure. The average values across the 15 participants were then calculated, with the results displayed in the figure below. The results show that the "decorative density" feature had the highest purity score at 0.77, while the other three features had purity values fluctuating around 0.7, indicating no significant difference among them. Regarding V-Measure, the "transparency degree" feature performed the best with a score of 0.52. The next two spatial features scored 0.37 and 0.31, respectively, while the "color scheme" feature performed the worst, with a score of only 0.23.
\begin{figure}[h]
  \centering
  \includegraphics[width=\linewidth]{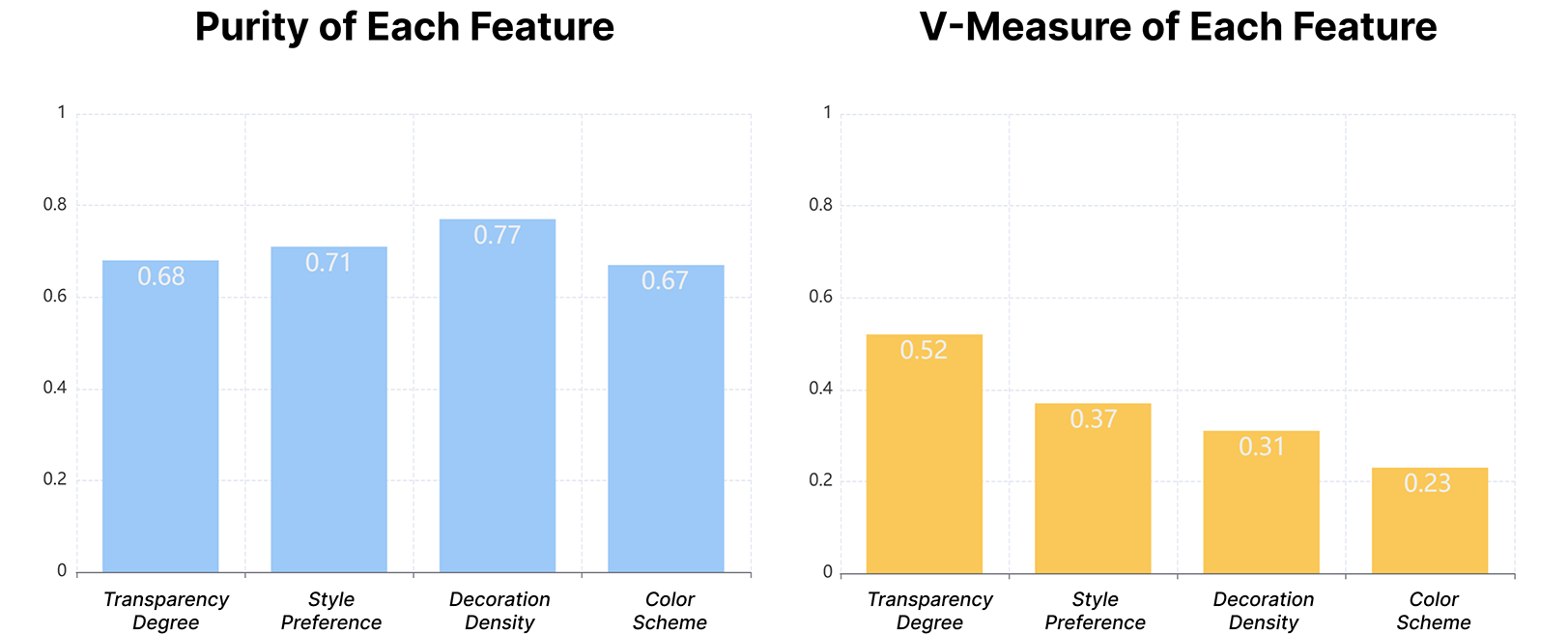}
  \caption{Evaluation of Clustering Results for Each Feature}
  \Description{Evaluation of Clustering Results for Each Feature}
\end{figure}

After evaluating the overall implementation completeness and prediction accuracy of the final brain-computer interaction design system, the researchers selected three commands from the aforementioned four spatial features, covering a total of eight spatial design directions. The selected commands for this study were: "increase transparency," "more luxurious decoration," and "more classical style." Objectively, these commands are more likely to induce similar brain activity patterns among participants, and from a practical standpoint, their combinations can yield a richer variety of spatial design outcomes.
\begin{figure}[h]
  \centering
  \includegraphics[width=\linewidth]{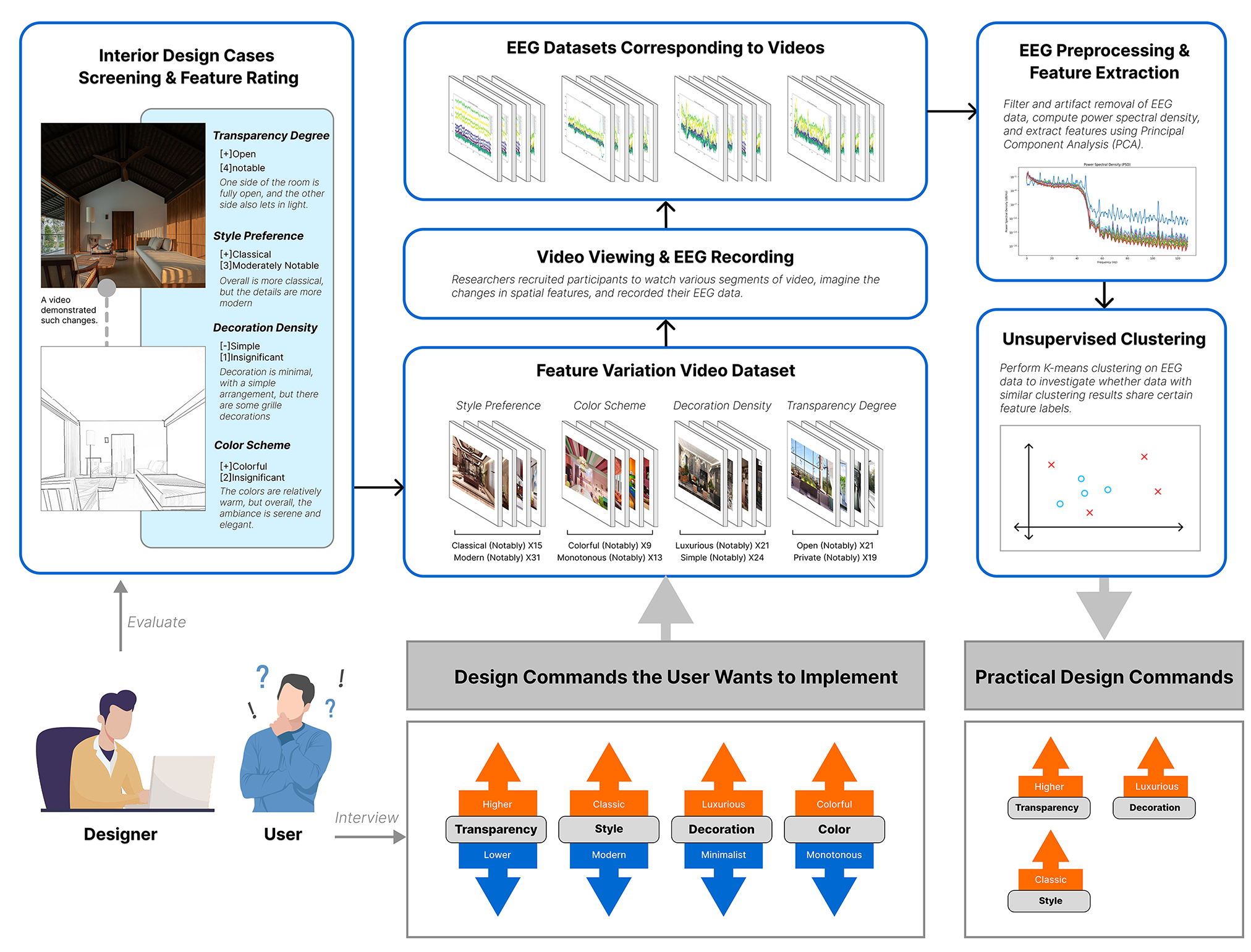}
  \caption{Selection Process for Design Instructions}
  \Description{Selection Process for Design Instructions}
\end{figure}

\section{PREDICTION METHODS}

\subsection{Generative AI Training}
The study utilized the Stable Diffusion platform to generate architectural space images for each spatial feature, which included tasks such as dataset creation, fine-tuning model training, prompt writing, and parameter adjustments. This step not only produced the materials required for subsequent training experiments but also provided assets for the brain-computer interface design system to utilize in response to EEG-based commands.

To enable users to express spatial design through imagined brain intent, the study first developed an image generation system for interior space design using the ComfyUI platform of Stable Diffusion. This system was designed to automatically generate spaces with specific features corresponding to the user’s intent. The research focused on optimizing two key aspects of generative artificial intelligence: prompt engineering and fine-tuning models (LoRA). For prompt engineering, the researchers expanded the selected spatial commands into detailed nouns and scene descriptions. After extensive testing, a set of effective prompts was developed to accurately realize the specified design commands. For fine-tuning models, the researchers scraped a substantial collection of raw images representing the target spatial features from the internet. These images were meticulously curated and refined using Photoshop for precise adjustments, creating distinct image datasets that correspond to the three specific spatial design commands. These datasets were then used to develop three corresponding LoRA fine-tuned models.

After completing the prompt construction and LoRA model fine-tuning, the generative AI system demonstrated a substantial capability to produce images with the desired spatial features. Additionally, when combined with the ControlNet model, the system gained the preliminary ability to modify input images in specific directions according to the user's intent. Based on unmodified images of raw interior spaces, the researchers used the trained tools to generate 93 images that showcased variations across the three spatial features: transparency degree, decorative density, and stylistic preference. Following an evaluation process, 60 images were selected for inclusion, evenly distributed across the three commands. These modified interior space images, paired with their original counterparts, were compiled to create a dataset of 60 image pairs.

\subsection{Assisted Video Construction}
After creating the image dataset, we utilized it to facilitate the training of brain-computer interface (BCI) commands. The typical process of training BCI commands involves the following steps: 1) Pre-demonstrating the execution process of the commands, usually in the form of tasks or stimuli; 2) Having participants observe the execution process, imagine performing the task, or receive the stimulus; 3) Collecting EEG data during the imagination phase to train a machine learning model; 4) Using the trained model to classify the corresponding imagined commands from the participant's EEG data, thereby enabling BCI. In this study, the image dataset was primarily used to guide participants in imagining the corresponding commands.

During the data collection phase of our experiment, to guide participants in imagining changes in architectural spaces, we used the unmodified raw interior image and the AI-generated image with specific features from each image pair as the first and last frames of a video, respectively. Various transformation techniques were applied between these two images to create the video. The research team considered how different video transformation methods might affect participants’ cognitive engagement, emotional involvement, and situational association during viewing. Based on scores from the Immersive Experience Questionnaire (IEQ), we ultimately selected the most straightforward "gradual transition" video generation algorithm to produce the guide videos for participant imagination. 

Using the image dataset, we generated 60 demonstration videos depicting the transformation of interior space features. Each video was 12 seconds long, containing three repetitions of the transformation process. Each transformation cycle included 1 second of still frames for both the initial and final images and a 2-second gradual transition animation. A 3-second black still frame was inserted between videos to enhance the participants’ viewing and imagination experience. The entire video viewing and command imagination experiment lasted for 15 minutes.

\subsection{Experiment Procedures}
In the main experimental procedure, we recruited a total of 7 participants, all of whom were university students. Among them, 3 were enrolled in a design school, while the other 4 had no design-related experience. 

Participants watched 60 sets of feature transformation videos and were informed that the architectural space design in the videos primarily aimed to alter one of three features: "transparency degree," "decorative density," or "style preference." Participants were required to identify the change in spatial features in the video and imagine that this change was occurring according to their own intentions. EEG data were collected from participants during the experiment, and a separate Support Vector Machine (SVM) model was trained for each participant using the feature labels. This was done to predict design intentions from EEG signals and compare the accuracy of these predictions. The EEG recording equipment used was the same as that described in section 3.2.
\begin{figure}[h]
  \centering
  \includegraphics[width=\linewidth]{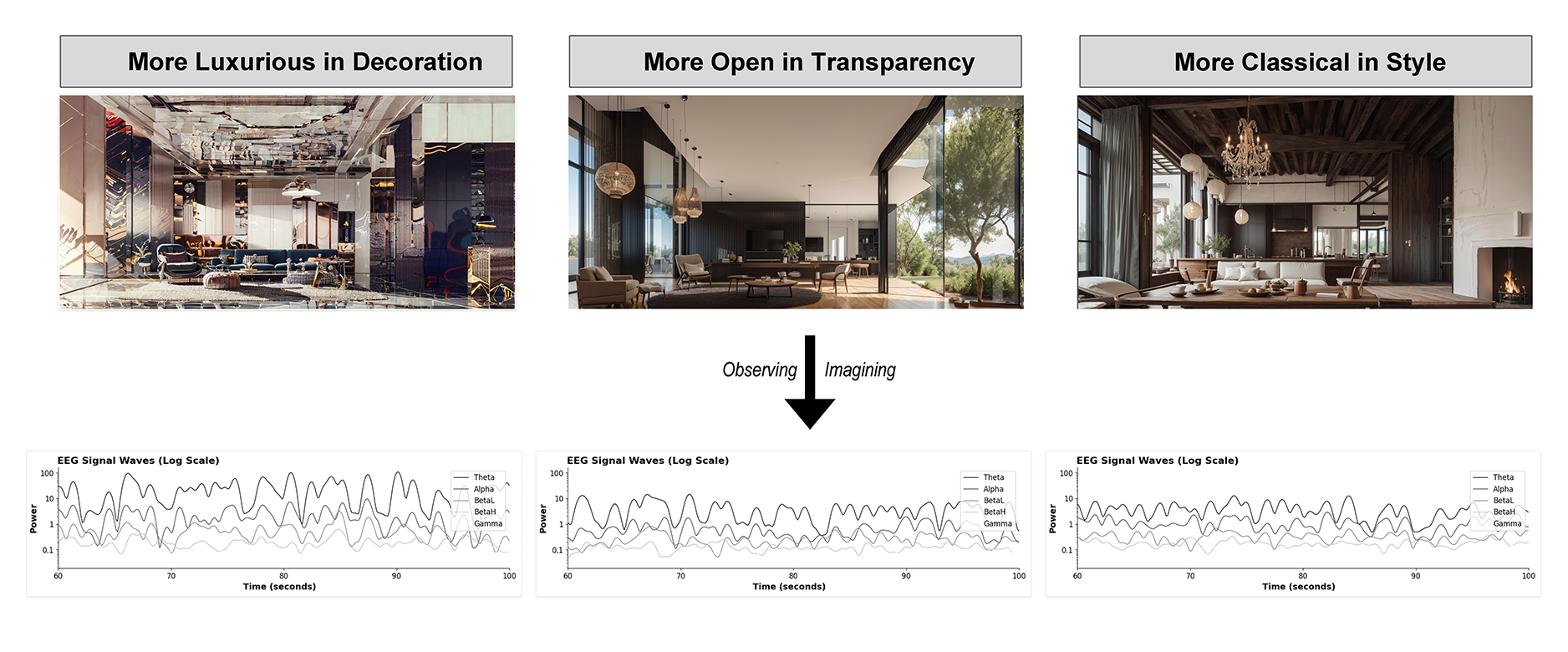}
  \caption{Corresponding Results for the Three Design Instructions}
  \Description{Corresponding EEG Results for the Three Design Instructions}
\end{figure}

During the process of collecting EEG data from participants' imagined tasks and training the machine learning models, participants were instructed to remain in the experimental space, maintain a meditative state, and stay relaxed. They were also encouraged to observe the surrounding space. In practice, this process typically lasted for about 30 minutes. Once the machine learning models used for classifying and predicting EEG commands were trained, participants experienced a session where they could use the trained model to engage in interactive design.

\subsection{SVM Model Training}
The researchers began by preprocessing the collected EEG data using a band-pass filter to remove power frequency noise and retain brainwaves within the 0.5Hz to 45Hz range. They then employed Independent Component Analysis (ICA) to isolate artifacts such as eye movements and muscle activity from the raw EEG signals. The EEG data was segmented into samples using a 2-second window with a 0.5-second overlap to ensure temporal continuity and comprehensive coverage of the brain activity. To ensure balanced contributions of each feature to the decision boundaries and to enhance the model's accuracy and generalizability, the researchers applied Z-score normalization. This standardization technique scaled the original data features to have a mean of 0 and a standard deviation of 1.

Using the preprocessed EEG data, a machine learning model was trained for each participant. For the three design commands and their corresponding EEG imagery data, the researchers utilized a ten-fold cross-validation method to train Support Vector Machine (SVM) models, aiming to classify and predict design intentions based on brain signals. SVMs work by finding the optimal hyperplane that maximizes the margin between classes, effectively establishing decision boundaries between different predictions. Given that a basic SVM is a binary classifier and this study involved three distinct spatial design commands, a one-vs-rest strategy was adopted. This approach involved training a separate classifier for each category, where the current feature data point was treated as the positive class and all other feature data points as the negative class. During prediction, the output was determined by the classifier with the highest confidence score. Additionally, prior research has demonstrated that linear SVMs, polynomial kernel SVMs, and Radial Basis Function (RBF) kernel SVMs achieve high accuracy in EEG data classification tasks. This study employed the RBF kernel for the SVM to enhance prediction accuracy.

After EEG data collection and SVM model training for all participants, the average prediction accuracy across different participants was calculated to represent the final outcome. The trained SVM models achieved an average prediction accuracy of approximately 68.67\% for user design commands. Specifically, the prediction accuracy was 0.61 for the "more classical style" command, 0.67 for the "more luxurious decoration" command, and 0.78 for the "increase transparency" command.

In summary, the effectiveness of model training for each command was positively correlated with the semantic clarity of the command in spatial cognition. For example, the "increase transparency" command, which suggests a more open spatial structure with a higher window-to-wall ratio and broader outdoor views from the interior, has a clear semantic meaning. In contrast, "increase decorative density" can be interpreted as enhancing the sophistication of interior decorations, adding texture changes to ceilings and wall surfaces, etc., resulting in increased visual information density in the space. However, the specific design techniques for achieving this are not always well-defined. The "style" feature covers a broad range; while people generally have basic concepts of modern versus traditional architectural styles, designing solely based on stylistic features tends to lack clear semantic guidance and can be challenging to imagine. The specific implementation of style also varies greatly depending on the context, spatial structure, and even the designer's personal style. Consequently, the SVM's accuracy in learning and predicting the three selected commands from the study ranked as follows: "increase transparency" > "more luxurious decoration" > "more classical style."

\section{MENTAL-GEN SYSTEM}
By applying machine learning techniques to the EEG data collected from each participant's design imagination process, the researchers developed a personalized SVM prediction model for each participant. These models, in conjunction with pre-trained generative AI models, were utilized to construct a brain-computer interface (BCI) system for spatial design tailored to each individual. Within this interactive design system, users provide real images of interior spaces that require design interventions and then imagine their desired design intentions. The system predicts and generates visual representations of the imagined design intentions based on the EEG data captured in each session. The users then evaluate whether the generated images align with their envisioned designs.
\begin{figure}[h]
  \centering
  \includegraphics[width=\linewidth]{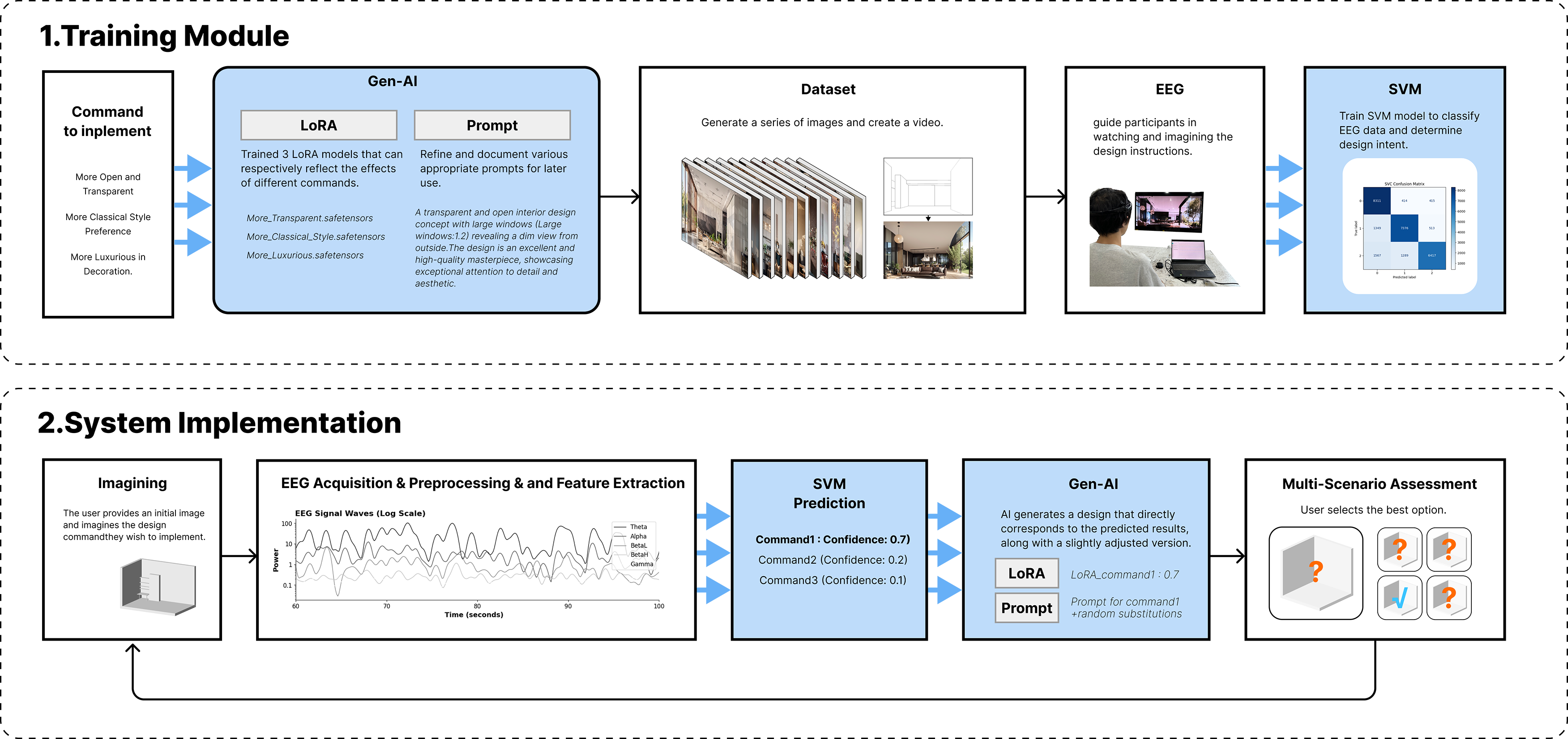}
  \caption{Training, Prediction, and Representation of Design Instructions}
  \Description{Training, Prediction, and Representation of Design Instructions}
\end{figure}

For system implementation, the study employed the Emotiv EPOC X device for EEG data acquisition and utilized the Openvibe platform for real-time EEG signal transmission. The pre-trained machine learning models then predict the design intentions from the EEG signals, and the results, along with their associated weights, are transmitted in real time via a WebSocket client to the Comfy UI workflow of the Stable Diffusion platform, where generative design of spatial images is performed.

\subsection{Design Command Extraction}
Initially, users observe an image of the initial space and imagine the desired changes they want to implement in the interior environment. During this process, some users may already have a fully developed and comprehensive design concept for the interior space, such as "a refined New Chinese-style home with dark wood flooring and mahogany furniture, where windows are discreetly designed, and suitable green plants are added to break up horizontal divisions." Alternatively, they might envision "a minimalist modern style interior with a cool color palette of black, white, and grey, using grey-black as a base tone with occasional bright accents, and featuring large floor-to-ceiling windows to create an expansive and open space." Meanwhile, some users may not have a complete design concept yet and instead may lean towards more fragmented and conceptual requirements, such as a "Nordic-style home," "a more minimalist design," or "a pet- and child-friendly space."

Given the complexity of the dimensions involved in architectural design, it is challenging to exhaustively enumerate and categorize all possible user design concepts. However, due to the interrelationships and mutual influences among different spatial design dimensions, which together determine the final design outcome, this study focuses on three specific spatial characteristics: "transparency degree," "decorative density," and "stylistic preference." Users are encouraged to concretize their existing design ideas along these three dimensions. For each session, users imagine one of these design intentions, which are then classified and predicted by the machine learning models. Through multiple iterations, the system gradually generates an image that represents the user's ideal design solution.

\subsection{System Implementation}
After the machine learning model completes the classification and prediction of the user's design intention based on EEG signals, the system identifies the spatial feature the user intends to modify and uses the confidence level of the classification as an indicator of the user's desire for change. Based on the classification results, the system selects the corresponding prompts and fine-tuned models trained in advance on the Stable Diffusion platform. The classification confidence level is then used as the weight for the fine-tuned model, which, together with the selected prompts, is utilized to regenerate the input image. During the generation process, the researchers also employed ControlNet models to analyze the user's original residential space images, primarily using Canny edge detectors and MLSD line segment detectors to extract the basic spatial framework and contours. This ensures that the final design scheme aligns with the original spatial framework, allowing it to be applicable in the user's real interior design context.

Within the BCI-driven interactive design system's interface, in addition to generating the most likely result as predicted by the machine learning model, the system also produces four additional spatial design images that do not fully align with the predicted outcome. These images are generated by partially randomizing, replacing, and regenerating the prompts corresponding to the predicted results. The replacement content is drawn from an interior design corpus accumulated by the research team during the exploration of prompts in Section 4.1, thereby broadening the user's choices and avoiding potential mismatches between the prompts and models fine-tuned by the researchers and the user's imagination. Once the five spatial design images are output, the user will proceed through the following three steps:

\begin{enumerate}
    \item The user compares the five output images to determine whether any of them fully realize their design vision. If so, they mark the image and conclude the BCI drawing process.
    
    \item If none of the five images from the current round fully satisfies the user, the user rates each image on a scale of 1 to 7, selecting the image they find most satisfactory overall. Satisfaction is measured by two criteria: "Does the generated image effectively realize the design intention I imagined?" and "Is the generated image a good interior design solution for me?"
    
    \item Using the selected image as the input, the system rereads the user's EEG-based design intention and iteratively redraws the spatial features that need adjustment, generating a new set of five design proposals.
\end{enumerate}

The study records the user’s satisfaction at each iteration and explores the relationship between satisfaction levels and the number of iterations.
\begin{figure}[h]
  \centering
  \includegraphics[width=\linewidth]{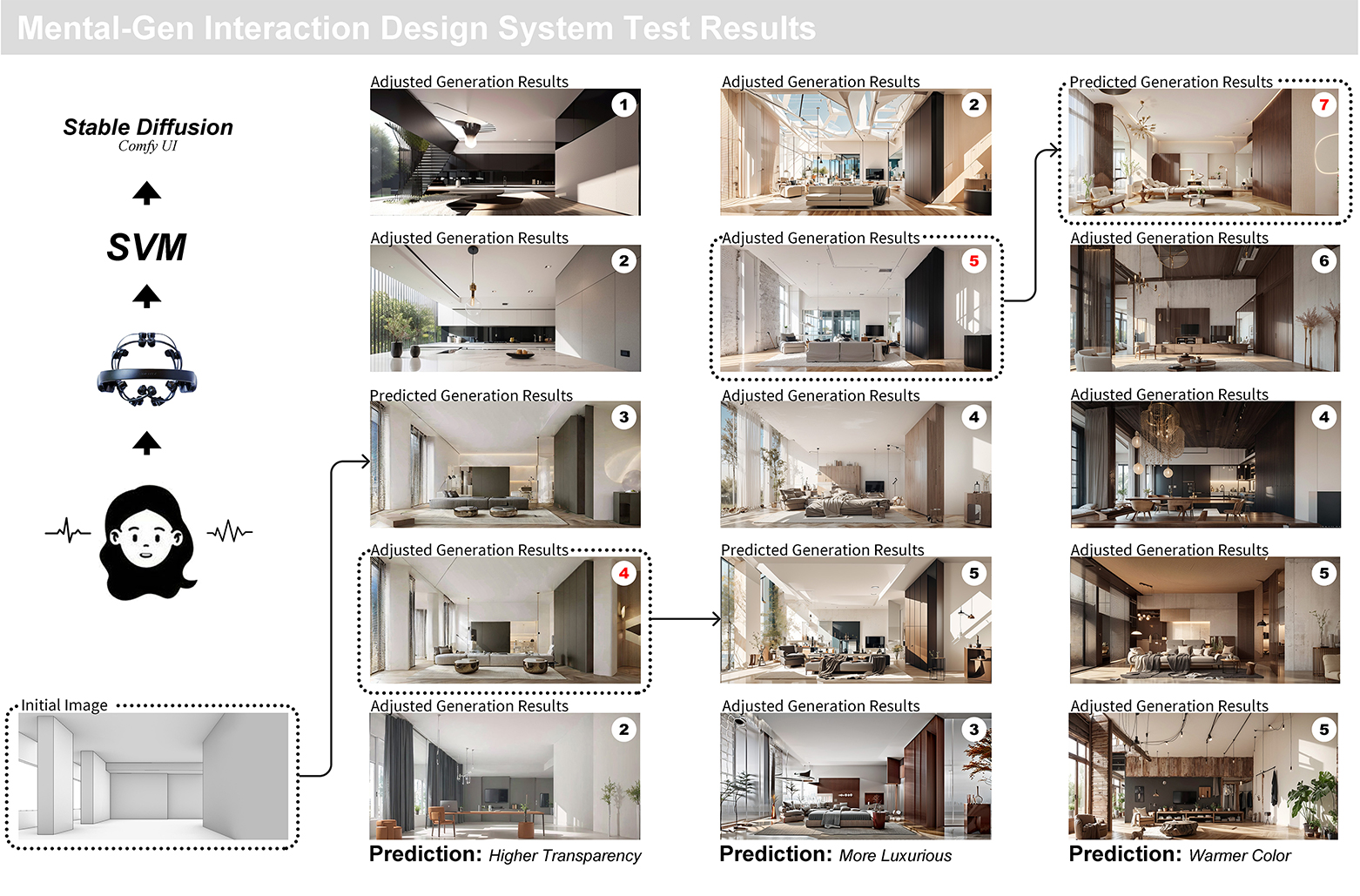}
  \caption{Example of System Testing Results}
  \Description{Example of System Testing Results}
\end{figure}

Following a waiting and relaxation period of approximately 30 minutes, participants from the experiment described in Section 4.3 proceeded to evaluate the BCI interactive design system. They were asked to select a base interior space image provided by the researchers and imagine their desired design modifications. The EEG acquisition device transmitted the participants' EEG signals during their imaginative process to the Openvibe platform, where the signals underwent preprocessing. The EEG data were then classified by the participant’s machine learning model to determine the design intention, which was finally transmitted via WebSocket to the local ComfyUI image generation system. The existing design scheme was modified according to the instructions and confidence levels indicated by the participant. Participants evaluated the five updated designs displayed on the screen, rating each on a scale from 1 to 7 for satisfaction. The researchers then used the highest-scoring design as the basis for guiding the participant into the next round of design imagination. Each participant engaged in at least eight rounds of intention imagining and design formulation. At the end of the experiment, each participant was also asked to report the accuracy and usability of the three types of design instructions they experienced.

The researchers documented each participant's satisfaction scores during the iterative BCI drawing process and conducted post-experiment interviews. The interviews focused on the following three aspects: "Did the system effectively recognize the design ideas in your mind?", "Did the system help you explore your design preferences?", and "What aspects of the system do you think could be improved?"

\subsection{System Evaluation}
During the final application of the brain-computer interface (BCI) design system, participants attempted to use the interactive system for interior space design. Their evaluation of the overall system and the three specific commands is reported as follows.

\textbf{Overall System Evaluation:} Participants' overall assessment of the system's predictive accuracy was approximately 4 out of 7 points (57.14\%), slightly lower than the predictive accuracy achieved during cross-validation of the model. The accuracy showed a slight increase followed by a decrease with each iteration of the commands, but overall remained relatively stable. In contrast, participants' reported satisfaction demonstrated a steady upward trend. This increase in satisfaction can be attributed to the iterative nature of the design process; since the initial input was a white rendering base map and each iteration modified only one spatial feature, the satisfaction scores for the initial outputs were generally low. The researchers believe that this was due to the artificial intelligence model's limited ability to transform highly abstract images lacking in detail. However, as design commands were iteratively applied and accumulated, the visual outcomes were gradually optimized, leading to an increase in participant satisfaction. The average satisfaction score for the first output was 2.4 out of 7, for the second output it was 3.1 out of 7, and for the third output, it was 2.9 out of 7. With later iterative generations, user satisfaction scores gradually increased and stabilized. By the seventh output, the average satisfaction score had reached 5.4 out of 7, indicating that the BCI design system at this point was largely capable of interpreting and expressing the user's design intentions based on EEG data, thus demonstrating a certain level of applicability.

\textbf{Specific Commands Evaluation:} Regarding the three specific commands, the accuracy reported by participants was also slightly lower than the values reported by the machine learning model but remained within an acceptable range (57.14\%-71.43\%). The command with the highest accuracy as reported by users was "Increase in Transparency," with a score of 5 out of 7. The overall satisfaction reported by users was slightly higher, which the researchers speculate may be due to the system's ability to produce good visual results even when the model's predictions deviated in the later stages. The command "Increase in Decoration Luxury," which involves increasing the complexity of space decoration, received the most polarized scores but also achieved the highest individual score. Based on the interviews conducted, we hypothesize that users prefer BCI commands that can lead to richer and more diverse design outcomes compared to those with more straightforward directives.
\begin{figure}[h]
  \centering
  \includegraphics[width=\linewidth]{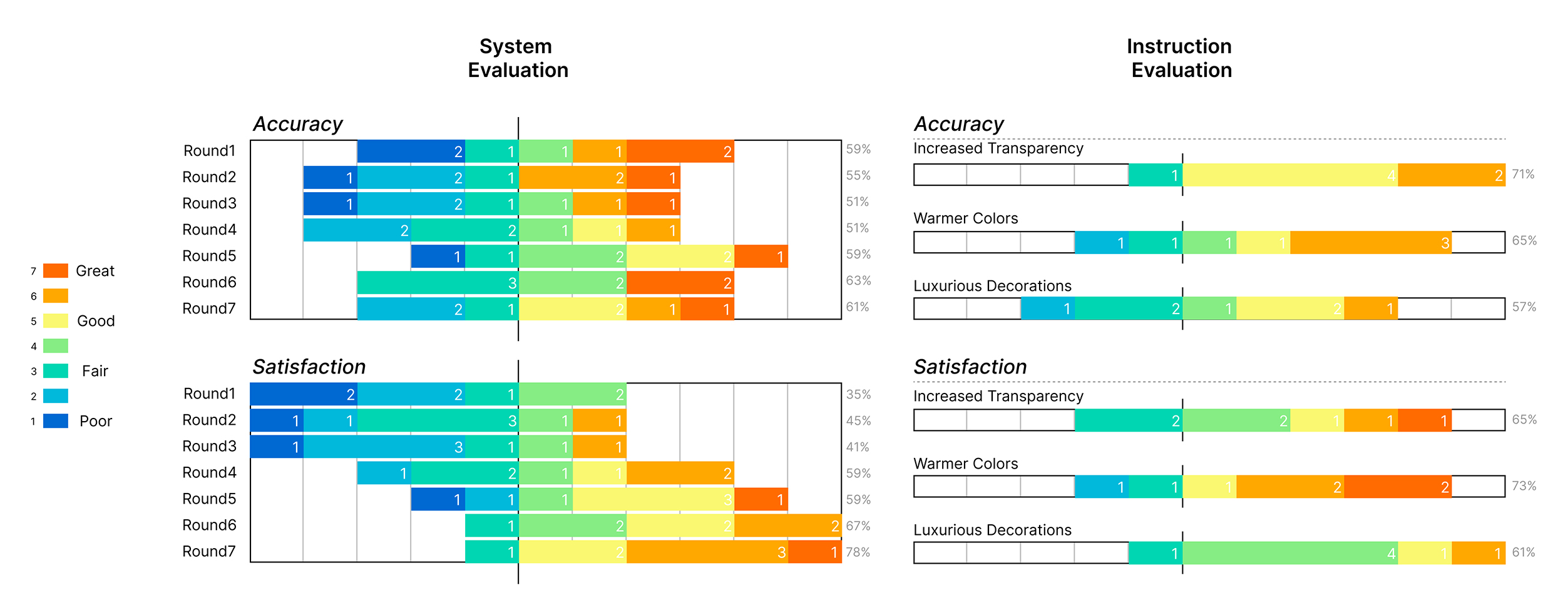}
  \caption{Analysis of User Feedback}
  \Description{Analysis of User Feedback}
\end{figure}

\section{DISCUSSION AND CONCLUSION}
During the learning process of EEG-based imagination for various spatial features, the researchers observed that commands associated with spatial features that are semantically clearer and more directly translatable to practical design actions demonstrated better machine learning performance. Similarly, EEG data corresponding to such spatial features exhibited better clustering density and purity in previous unsupervised clustering analyses. Furthermore, during the unsupervised clustering of EEG data, researchers found that clusters with the strongest density corresponded to specific interior design images, revealing that participants exhibited more consistent EEG responses to design changes in certain spatial orientations. These findings suggest areas for further exploration regarding the mechanisms of EEG response to diverse spatial characteristics.

In the post-trial interviews regarding user experience with the final system, researchers received an average response indicating that the interactive design system was "generally satisfactory." Some users provided suggestions for the selected spatial features, such as: "From the perspective of brain-computer interaction, having the option to open windows is quite feasible, but such design ideas are relatively rare," and "The ability to adjust spatial complexity is very practical." Additionally, since this study defined the style preference command as a transition from modern to traditional Chinese interior styles, a few users mentioned, "Sometimes I would like to change the interior style, but the results generated after EEG recognition diverged from what I wanted." This highlights the need for further exploration of EEG response mechanisms for more diverse and mixed spatial characteristics. Furthermore, since the final experiment involved students, some participants expressed in the interviews that "in subsequent iterations, the need for spatial changes felt less pressing, and the AI-generated results were quite satisfactory, so anything would work." Although this did not impact the experimental methodology, it suggests that future research should involve a broader range of real residential users to enhance the indoor space BCI design system developed in this study.

In summary, this study explores the potential for general users to express their design ideas using brain-computer interface (BCI) technology. Initially, the research investigated valuable spatial features for interior design and performed unsupervised clustering of users' EEG data associated with these features. It was found that EEG responses to the spatial features of “transparency,” “decorative density,” and “style preference” were the most consistent. Subsequently, generative artificial intelligence models were trained based on these spatial features, and spatial transformation videos were created to train a Support Vector Machine (SVM) model for predicting EEG responses related to design commands. The overall accuracy of the model training results was 68.67\%, with the command “Increase in Transparency” achieving the highest accuracy of 0.78. Ultimately, by integrating the pre-trained generative tools with the machine learning classification model, an interactive design system was developed. The system achieved an average user satisfaction score of 5.4 out of 7 in iterative generation results, efficiently meeting the goal of enabling users to design spaces directly through EEG imagination.

\bibliographystyle{ACM-Reference-Format}
\bibliography{sample-base}

\end{document}